\documentclass[twocolumn,showpacs,preprintnumbers,amsmath,amssymb,epsf]{revtex4}

\usepackage{graphicx}% Include figure files
\usepackage{dcolumn}% Align table columns on decimal point
\usepackage{bm}% bold math

\begin{document}

%preprint{APS/123-QED}

\title{Phase transitions in the $q$-voter model with two types of stochastic driving.}% Force line breaks with \\

\author{Piotr Nyczka, Katarzyna Sznajd-Weron, Jerzy Cis{\l}o}
%\email{kweron@ift.uni.wroc.pl} \homepage{http://www.ift.uni.wroc.pl/~kweron}
\affiliation{Institute of Theoretical Physics, University of Wroc{\l}aw, pl. Maxa
Borna 9, 50-204 Wroc{\l}aw, Poland }

\date{\today}

\begin{abstract}
In this paper we study nonlinear $q$-voter model with stochastic driving on a complete graph. We investigate two types of stochasticity that, using the language of social sciences, can be interpreted as different kinds of nonconformity. From a social point of view, it is very important to distinguish between two types nonconformity, so called anti-conformity and independence. A majority of works suggests that these social differences may be completely irrelevant in terms of microscopic modeling that uses tools of statistical physics and that both types of nonconformity play the role of so called 'social temperature'. In this paper we clarify the concept of 'social temperature' and show that different type of 'noise' may lead to qualitatively different emergent properties. In particularly, we show that in the model with anti-conformity the critical value of noise increases with parameter  $q$, whereas in the model with independence the critical value of noise decreases with the $q$. Moreover, in the model with anti-conformity the phase transition is continuous for any value of $q$, whereas in the model with independence the transition is continuous for $q \le 5$ and discontinuous for $q>5$.
\end{abstract}


\maketitle

\section{Introduction}
\label{intro}
Recently, various microscopic models of opinion dynamics has been proposed and widely studied  by physicists and social scientists (for reviews see \cite{CFL2009,RR2011,MW2002,GPL2003}). In a world of social studies this kind of modeling is known as the Agent-Based Modeling (ABM). It has been noted recently that despite the power of ABM in modeling complex social phenomena, widespread acceptance in the highest-level economic and social journals has been slow due to the lack of commonly accepted standards of how to use ABM rigorously \cite{RR2011,G2005}. As has been pointed out by Macy and Willer \cite{MW2002}, one of the main problems in the field of social simulations is \emph{'little effort to provide analysis of how results differ depending on the model designs'}. 

The similar problem is visible in a field of sociophysics. For example, to study opinion dynamics under conformity (one of the major paradigm of social response), a whole large class of models based on binary opinions $S = \pm 1$ has been proposed, among them voter model \cite{HL1975}, majority rule \cite{G2002,KR2003}, Sznajd model \cite{SWS2000} or nonlinear voter models \cite{LR2008,CMP2009}. For all these models the ferromagnetic state is an attractor \cite{CFL2009}. On one hand, this is expected since the conformity is the only factor influencing opinion dynamics in these models. On the other hand this is obviously not realistic for real social systems. To make models of opinion dynamics more realistic several modifications has been proposed, among them contrarians  \cite{G2004,LLW2005}, inflexibles \cite{GJ2007} and zealots \cite{M2003}. From the social point of view all these modifications describe another major paradigm of social response - so called nonconformity \cite{Nail_2000}. There are two widely recognized types of nonconformity: anti-conformity and independence. From a social point of view, it is very important to distinguish between these two types of nonconformity \cite{SBAH2006,Nail_2000}. The term `independence' implying the failure of attempted group influence.  Independent individuals evaluate situations independently of the group norm. From these point of view both zealots introduced by Mobilia \cite{M2003}, as well as inflexibles introduced by Galam \cite{GJ2007} describe particular type of independent behavior. On the contrary, anti-conformists are similar to conformers in the sense that both take cognizance of the group norm -- conformers agree with the norm, anticonformers disagree. Therefore, contrarians introduced by Galam in \cite{G2004} or stochastic driving proposed by Lama et. al. \cite{LLW2005} describe anti-conformity. 

Although differences between two types of nonconformity are very important for social scientists, the results obtained so far indicate that differences may be irrelevant from the physical point of view. Both, contrarian and independent behaviors, play the role of social temperature that induce an order--disorder transition \cite{G2004,LLW2005,STT2011,NSC2012}. However, addressing the problem posed by Macy and Willer \cite{MW2002} we would like to check rigorously the differences between two types of nonconformity under the framework of a possibly general model of opinion dynamics. In a class of models with binary opinions such a general model has been recently introduced in \cite{CMP2009} under the name of $q$-voter model. As special cases this model consists of both linear voter, as well as Sznajd model. In this paper we investigate this model in the presence of different types of nonconformity and check if results for anti-conformity and independence are qualitatively the same, according to our first expectation. It should be mentioned that another general class of opinion dynamics, known as a majority rule \cite{G2002,KR2003}, would also be a good candidate to test the differences between two types of nonconformity. However, introducing a general type of independence seem not so straightforward in this case.

The paper is organized as follows. In the next section we introduce the generalized model with two types of nonconformity on a complete graph (topology which is particularly convenient for analytical calculations). In Section \ref{time_evol} we analyze the time evolution of the system described by the master equation. In this section we will already see differences between models with anti-conformity and independence, on contrary to our first prediction. In Section \ref{stat} we calculate analytically the stationary values of public opinion in a case of infinite system. Results presented in Sections \ref{time_evol} and \ref{stat} indicate clearly a phase transition between phases with and  without majority. Therefore in the section \ref{crit} we find the phase diagram and calculate the transition point as a function of model's parameters. Results presented in this section show clearly important qualitative differences between two types of nonconformity. In the section \ref{sec_landau} we apply the approach that has been used to study non-equilibrium systems with two $(Z_2)$ - symmetric absorbing states in \cite{HCDM2005,VL2008} to understand more deeply these differences and in particularly the origin of the discontinuous phase transition in the case of nonconformity .
We conclude the paper in the last section.

\section{Model}
\label{mod}
We consider a set of $N$ individuals on a complete graph, which are described by the binary variables $S=\pm 1$. At each elementary time step $q$ individuals $S_1,\ldots,S_q$ (denoted by $\uparrow$ for $S_i=1$ or $\downarrow$ for $S_i=-1$, where $i=1,\ldots,q$) are picked at random and form a group of influence, lets call it $q$-lobby. Then the next individual ($\Uparrow$ or $\Downarrow$), on which the group can influence is randomly chosen, we call it voter. 

The part of a model described above is a special case of nonlinear $q$-voter model introduced in \cite{CMP2009}. In the original $q$-voter model, if all $q$ individuals are in the same state, the voter takes their opinion; if they do not have a unanimous opinion, still a voter can flip with probability $\epsilon$. For $q=2$ and $\epsilon=0$ the model is almost identical with Sznajd model on a complete graph \cite{SL2003}. The only difference is that in the $q$-voter repetitions in choosing neighbors are possible. In \cite{PST2011} $q$-voter model with $\epsilon=0$ and without repetition has been considered on a one-dimensional lattice. In this paper we also deal with a $q$-voter model with $\epsilon=0$ and without repetition, but additionally we introduce a certain type of noise to the model. Original voter model describes only conformity, whereas the noise is introduced to describe nonconformity. 

In our model conformity and anti-conformity take place only if the $q$-lobby is homogeneous i.e all $q$ individuals are in the same state. In a case of conformity voter takes the same decision as the $q$-lobby (like in the original $q$-voter model), whereas in a case of anti-conformity the opposite opinion to the group. In a case of independent behavior, voter does not follow the group, but acts independently -- with probability $1/2$ it flips to the opposite direction, i.e. $S_{q+1} \rightarrow -S_{q+1}$. 

To check the differences in results that are caused by different types of nonconformity we consider three versions of the model:
\begin{itemize}
\item
\textbf{Anti-conformity I} - with probability $p_1$ voter behaves like conformist and with $p_2$ like anti-conformist. This type of anti-conformity has been investigated in a case of the Sznajd model on a complete graph in \cite{NSC2012}. Because results depend only on the ratio $p=p_1/p_2$, in this paper we consider $p_1=1$ and $p_2=p$. In this version of the model the following changes are possible:
\begin{eqnarray}
\underbrace{\uparrow\uparrow \ldots \uparrow}_{q} \Downarrow & \stackrel{p_1=1}{\longrightarrow} & \underbrace{\uparrow\uparrow \ldots \uparrow}_{q} \Uparrow \nonumber\\
\underbrace{\downarrow\downarrow \ldots \downarrow}_{q}\Uparrow & \stackrel{p_1=1}{\longrightarrow} & \underbrace{\downarrow\downarrow \ldots \downarrow}_{q}\Downarrow \nonumber\\ 
\underbrace{\uparrow\uparrow \ldots \uparrow}_{q} \Uparrow & \stackrel{p_2=p}{\longrightarrow} & \underbrace{\uparrow\uparrow \ldots \uparrow}_{q} \Downarrow \nonumber\\
\underbrace{\downarrow\downarrow \ldots \downarrow}_{q} \Downarrow & \stackrel{p_2=p}{\longrightarrow} & \underbrace{\downarrow\downarrow \ldots \downarrow}_{q} \Uparrow.
\label{def_mod_anty_1}
\end{eqnarray}
In other cases nothing changes.
\item
\textbf{Anti-conformity II} - with probability $p$ voter behaves like anti-conformist and with $1-p$ like conformist. This type of anti-conformity has been investigated in a case of the Sznajd model on several networks in \cite{LLW2005} and results were qualitatively the same as in \cite{NSC2012}. Indeed it is quite easy to notice that Anti-conformity II is a special case of Anti-conformity I. However, for the record we consider here both cases and show that indeed differences are only quantitative. In this case the following changes are possible:
\begin{eqnarray}
\underbrace{\uparrow\uparrow \ldots \uparrow}_{q} \Downarrow &\stackrel{1-p}{\longrightarrow}& \underbrace{\uparrow\uparrow \ldots \uparrow}_{q} \Uparrow \nonumber \\
\underbrace{\downarrow\downarrow \ldots \downarrow}_{q}\Uparrow &\stackrel{1-p}{\longrightarrow}& \underbrace{\downarrow\downarrow \ldots \downarrow}_{q}\Downarrow \nonumber \\
\underbrace{\uparrow\uparrow \ldots \uparrow}_{q} \Uparrow & \stackrel{p}{\longrightarrow} & \underbrace{\uparrow\uparrow \ldots \uparrow}_{q} \Downarrow \nonumber \\
\underbrace{\downarrow\downarrow \ldots \downarrow}_{q} \Downarrow & \stackrel{p}{\longrightarrow}& \underbrace{\downarrow\downarrow \ldots \downarrow}_{q} \Uparrow. 
\label{def_mod_anty_2}
\end{eqnarray}
In other cases nothing changes.
\item
\textbf{Independence} - with probability $p$ voter behaves independently and with $1-p$ like conformist. In a case of independent behavior individual changes to the opposite state with probability $1/2$. The following changes are possible:
\begin{eqnarray}
\underbrace{\uparrow\uparrow \ldots \uparrow}_{q} \Downarrow & \stackrel{1-p}{\longrightarrow} & \underbrace{\uparrow\uparrow \ldots \uparrow}_{q} \Uparrow\nonumber \\
\underbrace{\downarrow\downarrow \ldots \downarrow}_{q}\Uparrow & \stackrel{1-p}{\longrightarrow} & \underbrace{\downarrow\downarrow \ldots \downarrow}_{q}\Downarrow \nonumber \\
\underbrace{\ldots}_{q} \Downarrow & \stackrel{p/2}{\longrightarrow} & \underbrace{\ldots}_{q} \Uparrow \nonumber \\
\underbrace{\ldots}_{q} \Uparrow & \stackrel{p/2}{\longrightarrow} & \underbrace{\ldots}_{q} \Downarrow.  
\label{def_mod_non}
\end{eqnarray}
In other cases nothing changes.
\end{itemize}

\section{Time evolution}
\label{time_evol}
In a single time step $\Delta_t$, three events are possible -- the number of 'up' spins $N_{\uparrow}$ increases or decreases by $1$ or remains constant. Of course all three events can be rewritten for the number of 'down' spins $N_{\downarrow}$ as $N_{\uparrow}+N_{\downarrow}=N$. Also concentration $c = N_\uparrow/N$ of spins 'up' increases or decreases by $\Delta_N=1/N$ or remains constant:
\begin{eqnarray}
\gamma^+(c) & = & Prob\left\{c \rightarrow c+\Delta_N \right\} \nonumber\\
\gamma^-(c) & = & Prob\left\{c \rightarrow c-\Delta_N \right\} \nonumber\\
\gamma^0(c) & = & Prob\left\{c \rightarrow c\right\} = 1-\gamma^+(c)-\gamma^-(c).
\end{eqnarray}

The time evolution of the probability density function of $c$ is given by the master equation:
\begin{eqnarray}
\rho(c,t+\Delta_t) &=& \gamma^+(c-\Delta_N)\rho(c-\Delta_N,t) \nonumber\\
                &+& \gamma^-(c+\Delta_N)\rho(c+\Delta_N,t) \nonumber\\
                &+& [1-\gamma^+(c)-\gamma^-(c)] \rho(c,t).
\label{eq_evol}
\end{eqnarray}
Of course analogous formula can be written for $N_{\uparrow}$. The exact forms of probabilities $\gamma^+(c)=\gamma^+(N_{\uparrow})=\gamma^+$ and $\gamma^-(c)=\gamma^-(N_{\uparrow})=\gamma^-$ depend on the version of the model and for the finite system are the following:
\begin{itemize}
\item
Anti-conformity I:
\begin{eqnarray}
\gamma^+ & = & \frac{N_\downarrow \prod\limits_{i=1}^q (N_\uparrow - i + 1) + p\prod\limits_{i=1}^{q+1} (N_\downarrow - i + 1)}{\prod\limits_{i=1}^{q+1} (N - i + 1)} \nonumber \\
\gamma^- & = & \frac{N_\uparrow \prod\limits_{i=1}^q (N_\downarrow - i + 1) + p\prod\limits_{i=1}^{q+1} (N_\uparrow - i + 1)}{\prod\limits_{i=1}^{q+1} (N - i + 1)}
\label{gamma_finite_a1}
\end{eqnarray}
\item
Anti-conformity II:
\begin{eqnarray}
\gamma^+ & = & \frac{(1-p)N_\downarrow \prod\limits_{i=1}^q (N_\uparrow - i + 1) + p\prod\limits_{i=1}^{q+1} (N_\downarrow - i + 1)}{\prod\limits_{i=1}^{q+1} (N - i + 1)} \nonumber\\
\gamma^- & = & \frac{(1-p)N_\uparrow \prod\limits_{i=1}^q (N_\downarrow - i + 1) + p\prod\limits_{i=1}^{q+1} (N_\uparrow - i + 1)}{\prod\limits_{i=1}^{q+1} (N - i + 1)}
\label{gamma_finite_a2}
\end{eqnarray}
\item
Independence:
\begin{eqnarray}
\gamma^+ & = & \frac{(1-p)N_\downarrow \prod\limits_{i=1}^q (N_\uparrow - i + 1) }{\prod\limits_{i=1}^{q+1} (N - i + 1)} + \frac{pN_\downarrow}{2N} \nonumber\\
\gamma^- & = & \frac{(1-p)N_\uparrow \prod\limits_{i=1}^q (N_\downarrow - i + 1) }{\prod\limits_{i=1}^{q+1} (N - i + 1)} + \frac{pN_\uparrow}{2N}.
\label{gamma_finite_n}
\end{eqnarray}
\end{itemize}
 
Whereas for the infinite system the above formulas take much simpler forms:
\begin{itemize}
\item
Anti-conformity I:
\begin{eqnarray}
\gamma^+ & = & (1-c)c^q + p(1-c)^{q+1} \nonumber\\
\gamma^- & = & c(1-c)^q + pc^{q+1} 
\label{gamma_infinite_a1} 
\end{eqnarray}
\item
Anti-conformity II:
\begin{eqnarray}
\gamma^+ & = & (1-p)(1-c)c^q + p(1-c)^{q+1} \nonumber\\
\gamma^- & = & (1-p)c(1-c)^q + pc^{q+1}
\label{gamma_infinite_a2}   
\end{eqnarray}
\item
Independence:
\begin{eqnarray}
\gamma^+ & = & (1-p)(1-c)c^q + p(1-c)/2 \nonumber\\
\gamma^- & = & (1-p)c(1 - c)^q + pc/2 
\label{gamma_infinite_n}  
\end{eqnarray}
\end{itemize}

Solving analytically master equation (\ref{eq_evol}) is not an easy task, but exact formulas for $\gamma^+$ and $\gamma^-$ allow for a numerical solution of the equation. For an arbitrary initial state the system reaches the same steady state. In a case of the infinite system the probability density function is a sum of delta functions  $\rho_{st}(c) = \delta(c - c_1) +  \delta(c - c_2) +  \ldots + \delta(c - c_k)$, whereas for the finite system $\rho_{st}(c)$ has maxima (peaks) for the $c=c_j \; j=1,\ldots,k$ which are getting higher and more narrow with the system size, approaching deltas for the infinite system. The number of peaks $k$ and values $c_1,\ldots,c_k$ depend on the version of the model, as well as on the model's parameters $p$ and $q$. 

\begin{figure}
\begin{center}
\includegraphics[scale=0.6]{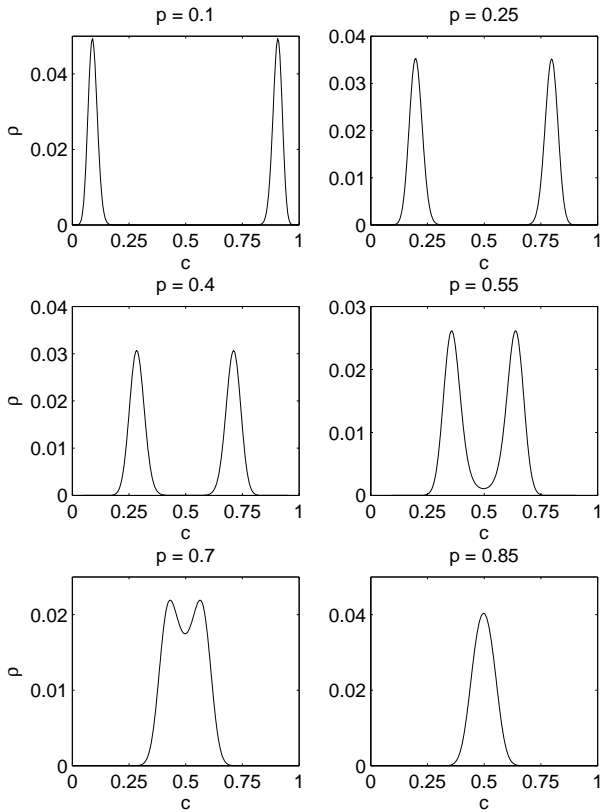}
\caption{Stationary probability density function of the concentration of 'up' spins for the $q$-voter model with anti-conformity I for the system of $N=200$ individuals and the size of the lobby $q=7$. As seen for small values of anti-conformity $p$ the system is polarized, but for large values of $p$ there is no majority in the system. For $p=0$ (the case without anti-conformity) the system consists of all spins 'up' or all spins 'down'. With increasing $p$ maxima are getting lower and approaching each other. Eventually they form a single maximum. This is typical behavior for a continuous phase transition. The critical value of $p$ can be found analytically (see section \ref{crit}) and depends on $q$.}
\label{pdf_aq7}
\end{center}
\end{figure}

\begin{figure}
\begin{center}
\includegraphics[scale=0.6]{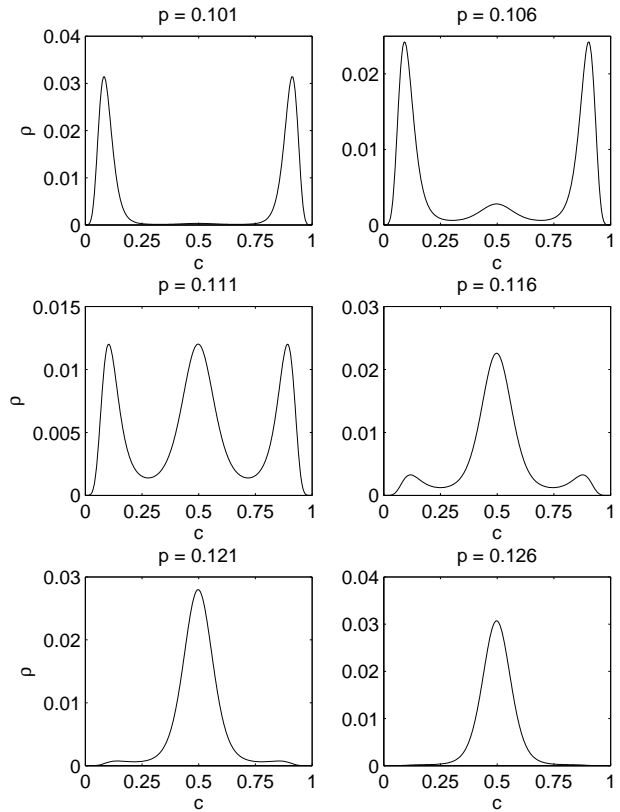}
\caption{Stationary probability density function of the concentration of 'up' spins for the $q$-voter model with independence for the system of $N=200$ individuals and the size of the lobby $q=7$. As seen for small values of independence $p$ the system is polarized, but for large values of $p$ there is no majority in the system. For $p=0$ (the case without independence) the system consists of all spins 'up' or all spins 'down'. For larger values of $p$ the third maximum appears at $c=1/2$ (no majority). This maximum increases with $p$ while the remaining two maxima are decreasing. Above certain value of $p$ there is only one maximum for $c=1/2$. This is typical behavior for a discontinuous phase transition for which we can observe the phase coexistence.}
\label{pdf_nq7}
\end{center}
\end{figure}

Examples of the stationary probability density functions for the $q$-lobby of size $q=7$ and the system size $N=200$ are presented in Figs. \ref{pdf_aq7} (anti-conformity I) and \ref{pdf_nq7} (nonconformity). As seen from Figs. \ref{pdf_aq7} and \ref{pdf_nq7} for small values of noise $p$ (whether the noise is introduced as independence or anti-conformity) the system is polarized, whereas for large values of $p$ there is no majority in the system. However, the transition from the phase with majority to the phase without majority is very different for each type of noise. In a case with anti-conformity we observe continuous phase transition for arbitrary value of $q$, whereas in a case with nonconformity there is a continuous phase transition for $q \le 5$ and discontinuous phase transition for $q>5$.

In the case with anti-conformity two states with majority, represented by two equally high peaks, are stable below the critical value of $p^*$. As $p<p^*$ increases the two peaks approach each other and  eventually for $p=p^*$ they form a single peak, which is a typical picture for a continuous phase transition (see Fig. \ref{pdf_aq7}) \cite{Callen1985,PB1989}. In a case with independence this picture is valid only for the lobby $q \le 5$. For $q>5$ the transition is very different. Again for small values of $p$ there are two peaks but with increasing $p$ they are not approaching each other. Instead, for $p=p_1^*$ the third peak appears at $c=1/2$ (see Fig. \ref{pdf_nq7}). The third peak is initially lower than remaining two peaks, which means that it represents a metastable state. As $p>p_1^*$ increases the third peak grows and for $p=p_2^*$ all three peaks have the same hight. For $p>p_2^*$ the central peak dominates over the other two, which means that the state $c=1/2$ is stable and remaining two are metastable. Finally, for $p=p_3^*$ side peaks disappear and only the center peak remains. This is a typical picture for a discontinuous phase transition, which takes place at $p=p_2^*$ \cite{Callen1985,PB1989}. Two values of independence parameter $p=p_1^*$ and $p=p_3^*$ demarcate the existence of metastability (spinodal lines) \cite{HHL2008,LB2002}. Values $p_1^*,p_2^*,p_3^*$ depend on the size of the lobby $q$, which will be shown exactly in the section \ref{crit}.

\begin{figure}
\begin{center}
\includegraphics[scale=0.6]{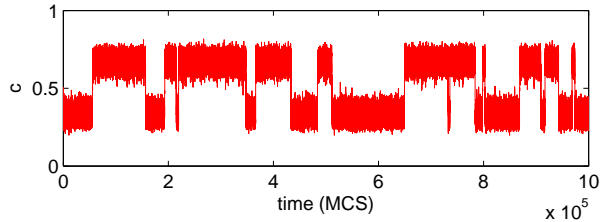}
\caption{Time evolution of the concentration of 'up' spins for the model with anti-conformity with lobby $q=7$ and level of anti-conformity $p=0.5$. The system size $N=200$. Spontaneous transitions between two stationary states are visible.}
\label{traj1}
\end{center}
\end{figure}

\begin{figure}
\begin{center}
\includegraphics[scale=0.6]{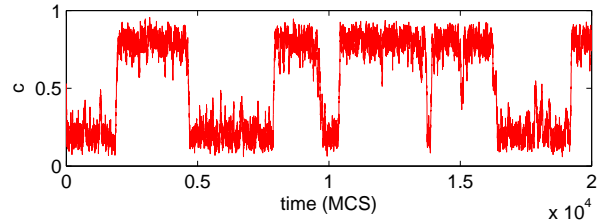}
\caption{Time evolution of the concentration of 'up' spins for the model with independence with lobby $q=5$ and level of anti-conformity $p=0.175$. The system size $N=200$. Spontaneous transitions between two stationary states are visible.}
\label{traj2}
\end{center}
\end{figure}

\begin{figure}
\begin{center}
\includegraphics[scale=0.6]{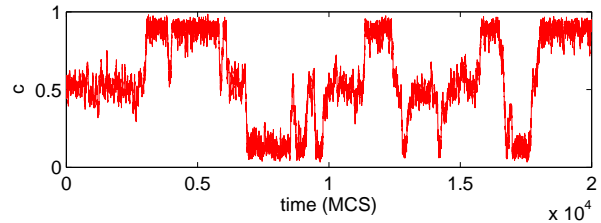}
\caption{Time evolution of the concentration of 'up' spins for the model with independence with lobby $q=7$ and level of anti-conformity $p=0.111$ (for this value all three states are stable). The system size $N=200$. Spontaneous transitions between three stationary states are visible.}
\label{traj3}
\end{center}
\end{figure}

\begin{figure}
\begin{center}
\includegraphics[scale=0.6]{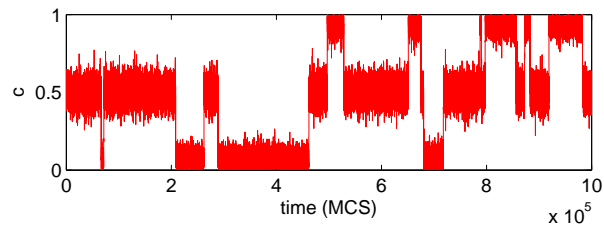}
\caption{Time evolution of the concentration of 'up' spins for the model with independence with lobby $q=9$ and level of anti-conformity $p=0.0685$. The system size $N=200$. Spontaneous transitions between three stationary states are visible.}
\label{traj4}
\end{center}
\end{figure}

Before moving on to the analytical results for the infinite system and determine the points of phase transitions, let us present the time evolution. We stop for a moment on the case of a finite system. Having exact formulas for transition probabilities $\gamma^+$ and $\gamma^-$ we are able not only to calculate numerically stationary density function $\rho_{st}(c)$ but also generate sample trajectories of concentration (Figs.\ref{traj1}, \ref{traj2}, \ref{traj3} and \ref{traj4}). In a case of a finite system spontaneous transitions between states are possible. In a case with anti-conformity transitions between two states, which correspond to peaks in the probability density function $\rho_{st}(c)$,  are possible below critical value of $p$. Because both peaks are equally high the system spends the same time on average in each state. 

This is also true in the case with independence and $q \le 5$ (see Fig. \ref{traj2}). However, as we have already written for $q>5$ there is a discontinuous phase transition between state with and without majority and for $p \in (p_1^*,p_3^*)$ there are three possible states. Therefore for $q>5$ we expect spontaneous transitions between three states.

Such transitions are indeed observed. For $p \in (p_1^*,p_2^*)$ state with majority is stable and state without majority is metastable. Therefore, the system spends more time in states with majority. For $p \in (p_2^*,p_3^*)$ the situation is exactly the opposite -- the state without majority is stable. In a transition point $p=p_2^*$ all three states are stable and the system spends the same time on average in each of three state (see Figs. \ref{traj3} and \ref{traj4}). 

\section{Stationary concentration}
\label{stat}
In the stationary state we expect that the probability of growth $\gamma^+$ should be equal to the probability of loss $\gamma^-$ and therefore:
\begin{equation}
F(c,q,p)=\gamma^+(c,q,p)-\gamma^-(c,q,p)=0, 
\end{equation}
where $F(c,q,p)$ can be treated as an effective force -- $\gamma^+$ drives the system to the state 'spins up', while $\gamma^-$ to 'spins down'. 
Therefore we can easily calculate also an effective potential:
\begin{eqnarray}
V(c,q,p) = - \int F(c,q,p) dc. 
\label{pot} 
\end{eqnarray}
To calculate stationary values of concentration we simply solve the equation:
\begin{equation}
F(c,q,p)=0,
\label{F0}
\end{equation}
or alternatively find minima of the potential $V$. Although the first possibility is more straightforward, we will see in the next section that knowing the form of potential will help us to calculate the transition points. 

Exact forms of the force $F$ and the potential $V$ for the infinite system are given below:
\begin{itemize}
\item
\textbf{Anticonformity I}
\begin{eqnarray}
F & = & (1-c)c^q + p(1-c)^{q+1} - c(1-c)^q - pc^{q+1} \nonumber \\
V & = & -\frac{1}{q+1} \left(c^{q+1} + (1-c)^{q+1}\right)\nonumber \\
& + & \frac{p+1}{q+2} \left(c^{q+2} + (1-c)^{q+2}\right)
\label{pot_a1}
\end{eqnarray}
\item
\textbf{Anticonformity II}
\begin{eqnarray}
F& = & (1-p)(1-c)c^q + p(1-c)^{q+1} \nonumber \\
& - & (1-p)c(1-c)^q - pc^{q+1} \nonumber \\
V& = & -\frac{1-p}{q+1} \left(c^{q+1} + (1-c)^{q+1}\right)\nonumber \\
& + & \frac{1}{q+2} \left(c^{q+2} + (1-c)^{q+2}\right)
\label{pot_a2}
\end{eqnarray}
\item
\textbf{Independence}
\begin{eqnarray}
F& = & (1-p)(1-c)c^q + \frac{p(1-c)}{2} \nonumber \\
& - & (1-p)c(1 - c)^q - \frac{pc}{2} \nonumber \\
V& = & -\frac{1-p}{q+1} \left(c^{q+1} + (1-c)^{q+1}\right)\nonumber \\
& + & \frac{1-p}{q+2} \left(c^{q+2} + (1-c)^{q+2}\right)
-\frac{p}{2} c (1-c)
\label{pot_n}
\end{eqnarray}
\end{itemize}

Solving analytically Eq. (\ref{F0}), i.e. finding $c_{st}$ as a function of $p$ for arbitrary value of $q$ is impossible, but we can easily derive the opposite relations satisfying Eq. (\ref{F0}):
\begin{itemize}
\item
Anticonformity I
\begin{eqnarray}
p=\frac{c_{st}(1-c_{st})^q - (1-c_{st})c_{st}^q}{(1-c_{st})^{q+1} - c_{st}^{q+1}}
\label{cp_a1}
\end{eqnarray}
\item
Anticonformity II
\begin{eqnarray}
p=\frac{c_{st}(1-c_{st})^q - (1-c_{st})c_{st}^q }{ (1-c_{st})^{q+1} + c_{st}(1-c_{st})^q - (1-c_{st})c_{st}^q - c_{st}^{q+1} }
\label{cp_a2}
\end{eqnarray}
\item
Independence
\begin{eqnarray}
p= \frac{c_{st}(1-c_{st})^q - (1-c_{st})c_{st}^q }{(1-c_{st})/2 + c_{st}(1-c_{st})^q - (1-c_{st})c_{st}^q - c_{st}/2}
\label{cp_n}
\end{eqnarray}
\end{itemize} 

\begin{figure}
\begin{center}
\includegraphics[scale=0.6]{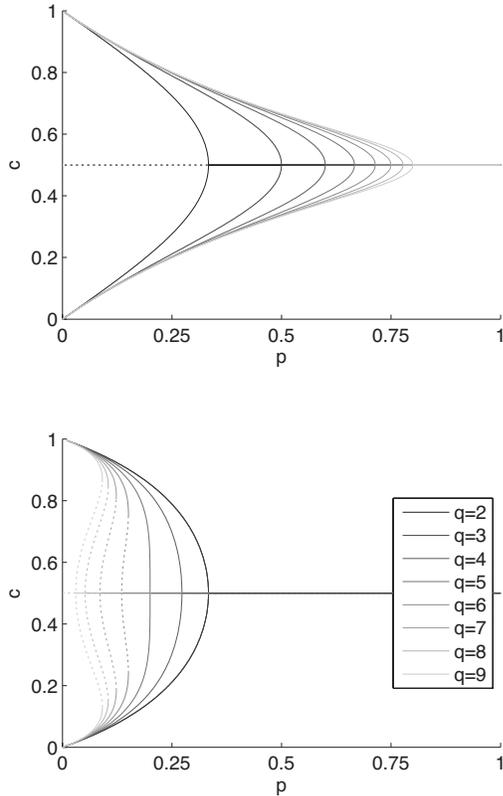}
\caption{Phase diagram for the model with anti-conformity (top panel) and independence (bottom panel). Dependencies between steady values of concentration $c_{st}$ and the level of noise $p$ for several values of $q$ are plotted using formulas (\ref{cp_a1})-(\ref{cp_n}). Although, only relation $p(c_{st})$ is calculated analytically and the opposite relation is unknown, we plot $c_{st}(p)$ simply rotating the figure. Dotted lines have been used to mark instability. Although both types of line (solid and dotted) are obtained from equation (\ref{F0}), i.e correspond to extreme values of potentials (\ref{pot_a1},\ref{pot_n}), only solid lines denotes stable values, i.e. correspond to the minima of potentials (see also Fig. \ref{flowdiagram}). The clear difference between two types of noise is visible -- in a case with anti-conformity the transition value of $p$ increases with $q$ and in a case with independence it decreases with $p$. Moreover, the type of transition is the same for arbitrary value of $q$ in a case with anti-conformity, whereas in a case with independence the transition between phases with and without majority changes its character for $q>5$.}
\label{cP}
\end{center}
\end{figure}

We have used the above formulas to plot the dependence between steady value of concentration $c_{st}$ and the level of noise $p$ for several values of $q$ (see Fig. \ref{cP}). Although, only relation $p(c_{st})$ is calculated analytically and the opposite relation is unknown, we plot $c_{st}(p)$ simply rotating the figure with the relation $p(c_{st})$. The clear differences between two types of noise are visible -- in a case with anti-conformity the transition value of $p$ increases with $q$ and in a case with independence it decreases with $p$. Moreover, the type of transition is the same for arbitrary value of $q$ in a case with anti-conformity, whereas in a case with independence the transition between phases with and without majority changes its character for $q>5$. 

\begin{figure}
\begin{center}
\includegraphics[scale=0.6]{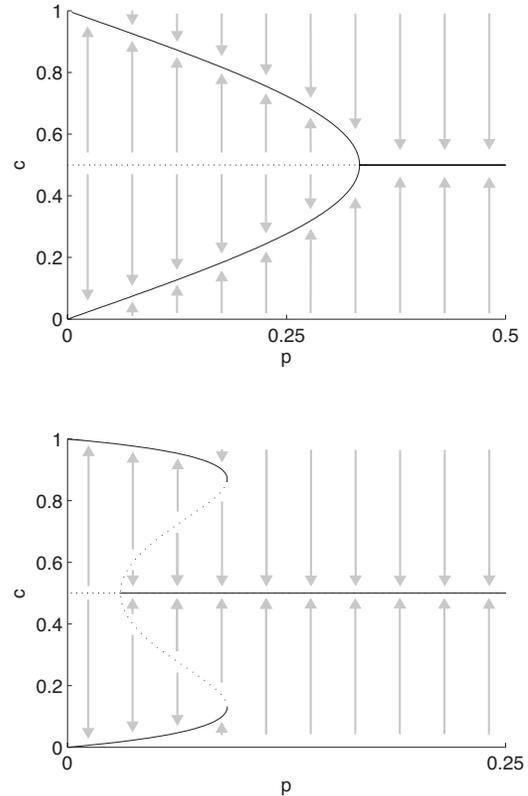}
\caption{Flow diagrams for the model with anti-conformity for $q=2$ (top panel) and independence for $q=9$ (bottom panel). Particular values of $q$ have been chosen just as examples and the dependences between stationary values of $c$ and parameter $p$ for other values of $q$ are seen in Fig. \ref{cP}. Here solid lines denote stable (attracting) steady values of concentration that correspond to the minima of potentials (\ref{pot_a1})--(\ref{pot_n}), whereas dotted lines denote unstable values of $c$ that correspond to maxima of potentials. Arrows denote the direction of flow, i.e. how the concentration changes in time. Particularly interesting behavior is related with independence (bottom) panel. Starting from two different initial concentrations disorder or order can be reached as a steady state (hysteresis).}
\label{flowdiagram}
\end{center}
\end{figure}

It should be also noticed that formulas (\ref{cp_a1})-(\ref{cp_n}) has been obtained from the condition (\ref{F0}), i.e correspond to extreme values of potentials (\ref{pot_a1})--(\ref{pot_n}). However, only minima of the potential correspond to the stable value of concentration. Therefore, in Figs. \ref{cP} and \ref{flowdiagram} we have denoted unstable values that correspond to the maxima of potentials by the dotted lines. Moreover, we have presented the flow diagram for chosen values of $q$ to show precisely which state is reached from a given initial conditions. Particularly interesting behavior is related with independence (bottom  panel in Fig. \ref{flowdiagram}). Starting from two  different initial concentrations disorder or order can be reached as a steady state (hysteresis). 

In the next section we derive analytically transition points using the knowledge of the effective potentials $V$ (\ref{pot_a1}), (\ref{pot_a2}) and (\ref{pot_n}).

\section{Phase transitions}
\label{crit}
As already noticed there is a continuous phase transition for the model with anti-conformity I and II for arbitrary value of $q$. Below critical value $p=p^*(q)$ an effective potential has two minima and above only one. Consequently stationary probability density function $\rho_{st}(c)$ for $p<p^*$ has two maxima and for $p>p^*$ only one at $c=1/2$ (there is no majority in the system). Analogous behavior is observed for the model with nonconformity but only for $q \le 5$. In all these cases we can easily calculate critical value $p^*$ making the simple observation concerning the behavior of the effective potentials (\ref{pot_a1}), (\ref{pot_a2}) and (\ref{pot_n}) for $q \le 5$ at $c=1/2$ (see also Fig. \ref{pdf_aq7} for clarity):
\begin{itemize}
\item
For $p<p^*$ potentials $V(c,p,q)$ have the maximum values for $c=1/2$ and therefore
\begin{eqnarray}
 \left. \frac{\partial^2 V(c,p,q)}{\partial c^2} \right|_{c=\frac{1}{2}} < 0
\end{eqnarray}
\item
For $p>p^*$ potentials $V(c,p,q)$ have the minimum values for $c=1/2$ and therefore
\begin{eqnarray}
\left. \frac{\partial^2 V(c,p,q)}{\partial c^2} \right|_{c=\frac{1}{2}} > 0.    
\end{eqnarray} 
\end{itemize}
This means that for  $p = p^*$ maximum changes to minimum at $c=1/2$:
\begin{eqnarray}
\left. \frac{\partial^2 V(c,p,q)}{\partial c^2}\right|_{c=\frac{1}{2}} = 0  \Rightarrow  \left.\frac{\partial F(c,p,q) }{\partial c}\right|_{c=\frac{1}{2}} = 0.   
\label{f12} 
\end{eqnarray}
Hence, the critical values:
\begin{itemize}
\item For anti-conformity I:
\begin{eqnarray}
p^*(q) = \frac{q - 1}{q + 1}
\end{eqnarray}
\item For anti-conformity II:
\begin{eqnarray}
p^*(q) = \frac{q - 1}{2q}
\label{eq_crit_anti2}
\end{eqnarray}
\item For independence with $q \le 5$:
\begin{eqnarray}
p^*(q) =  \frac{q - 1}{q - 1 + 2^{q-1}}
\label{eq_crit_indep}
\end{eqnarray}
\end{itemize}

\begin{figure}
\begin{center}
\includegraphics[scale=0.5]{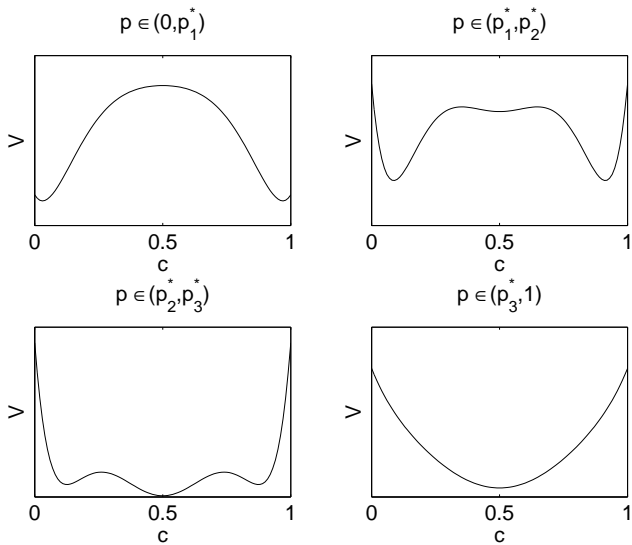}
\caption{Schematic plot of a potential for the model with independence and $q>5$. For $p \in (0,p_1^{*})$, potential $V(c,p,q)$ has $2$ minima that correspond to the states with majority. For $p \in (p_1^{*},p_2^{*})$ the potential $V(c,p,q)$ has $3$ minima and the state without majority is metastable. For $p \in (p_2^{*},p_3^{*})$ the potential $V(c,p,q)$ has $3$ minima and the states with majority are metastable. Finally, for $p \in (p_3^{*},1)$ the potential $V(c,p,q)$ has only $1$ minimum that corresponds to the state without majority. Exact form of the potential is given by Eq. (\ref{pot_n}).}
\label{fig_pot}
\end{center}
\end{figure}

As we see simple calculations allowed us to find critical points for almost all cases, except of the model with nonconformity for $q \ge 6$. In all cases considered above, there is a continuous phase transition between phases with and without majority. However, for model with independence and $q \ge 6$ the phase transition becomes discontinuous which has been already discussed in the section \ref{time_evol}. This behavior can be also suspected from the form of an effective potential (\ref{pot_n}) which for $q \ge 6$ has the following properties (see also Fig. \ref{fig_pot}):
\begin{itemize}
\item for $p \in \langle 0,p_1^{*})$, $V(c,p,q)$ has $2$ minima
\item for $p = p_1^{*}$ in $V(c,p,q)$ $3$-rd minimum emerges
\item for $p \in (p_1^{*},p_2^{*})$, $V(c,p,q)$ has $3$ minima
\item for $p = p_2^{*}$, $V(c,p,q)$ has $3$ minima which are equal
\item for $p \in (p_2^{*},p_3^{*})$, $V(c,p,q)$ has $3$ minima
\item for $p = p_3^{*}$ in $V(c,p,q)$  $3$-rd minimum disappears
\item for $p \in (p_3^{*},1 \rangle$, $V(c,p,q)$ has $1$ minimum  
\end{itemize}
As we see, there is an interval $ p \in (p_1^{*},p_3^{*})$ in which potential $V(c,p,q)$ has 3 minima and therefore stationary probability density function has $3$ maxima (see Fig. \ref{pdf_nq7}). In this region we have a coexistence of two phases - with and without majority. For $p<p_2^*$ the state with majority is stable and the state without majority is metastable and for $p>p_2^*$ the state with majority is metastable and without majority stable. Consequently the phase transition appears at $p=p_2^*=p^*$ and $p = p_1^{*},p_3^{*}$ designate spinodal lines \cite{HHL2008,LB2002}. 

\begin{figure}
\begin{center}
\includegraphics[scale=0.6]{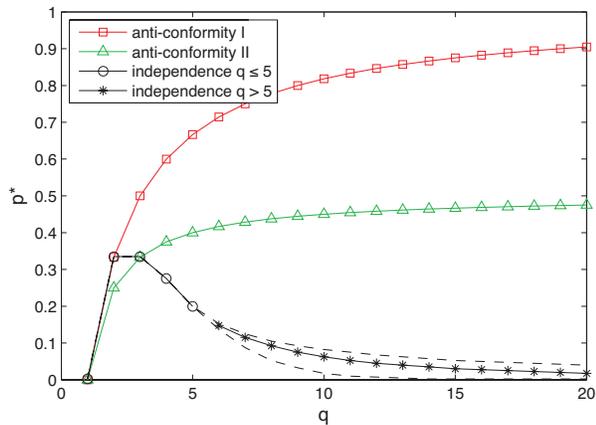}
\caption{Transition points $p^*$ as a functions of $q$ for all three models. Solid lines denote the line of the phase transition and dashed lines denotes spinodal lines i.e. determine the region with metastability. Several differences between models are visible. As seen both models with anti-conformity behave qualitatively the same - critical value of $p$ increases with $q$. However, for the model with independence the transition point decreases with $p$. Moreover, for $q \ge 6$  phase transition changes its type from continuous to discontinuous.}
\label{phase_tran}
\end{center}
\end{figure}

\begin{figure}
\begin{center}
\includegraphics[scale=0.6]{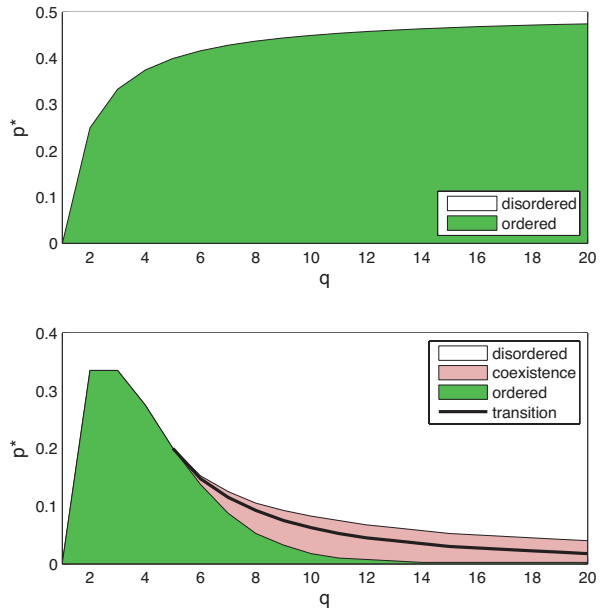}
\caption{Phase diagrams for the model with anti-confomity II (top panel) and independence (bottom panel).  As seen for the model with anti-conformity critical value $p^*$ increases and for the model with independence decreases with $q$. For anti-conformity (top panel) there is a continuous phase transition (denoted by the solid line) between order (i.e. $c \ne 1/2$ or equivalently $m \ne 0$) and disorder (i.e. $c = 1/2$ or equivalently $m = 0$). In the model with independence (bottom panel) there is a continuous phase transition only for $q<5$.  At $q=5$ the phase transition changes its type from continuous to discontinuous. For $q>5$ an area in which one of two phases (ordered or disordered) is metastable is limited by so called spinodal lines. This area is signed as 'coexistence', although the real coexistence occurs only on the transition line. However, in the region of metastability both phases can be observed depending on the initial conditions (hysteresis), which can be also seen from the flow diagram in Fig.\ref{flowdiagram}.}
\label{phase_tran1}
\end{center}
\end{figure}

Transition points $p^*$ as a functions of $q$ for all three models are presented in Figure \ref{phase_tran}. As seen, both models with anti-conformity (I and II) behave qualitatively the same - critical value of $p$ increases with $q$. However, for the model with independence the transition point decreases with $p$. Moreover, for $q=5$ in a case with independence, phase transition changes its type from continuous to discontinuous. To clarify our results we decided to present the complete phase diagrams for the models with anti-conformity and independence in Fig.\ref{phase_tran1}. Because results for both models with anti-conformity (I and II) are qualitatively the same we present the phase diagram only for the model with anti-conformity II.

The first difference between models with anti-conformity and independence connected with qualitative dependence between $p^*$ and $q$ is easy to explain heuristically. This is quite obvious why in the model with independence the critical point $p^*$ decreases when $q$ increases. When $q$ increases it becomes unlikely to choose randomly $q$ parallel spins and therefore the noise term dominates because it is independent on the state of the $q$-lobby.  Similarly, it can be understood why in the model with anti-conformity the critical point $p*$ increases with $q$. It should be recall here that anti-conformity takes place only when $q+1$ parallel spins are chosen randomly, which is more unlikely than choosing $q$ parallel spins.  Therefore the anti-conformity term decline in importance even more than conformity term as $q$ increases. The second difference between models that corresponds to the change of the transition type in the model with independence is not so easy to understand intuitively. This result has been obtain numerically from the potential (\ref{pot_n}), but in the next section we will show how this result could be also derived from an approximate Landau description.

\section{Landau description}
\label{sec_landau}
Although we were able to calculate critical points for the model with anti-conformity  and for the model with independence with $q \le 5$ directly from the potentials (\ref{pot_a1}), (\ref{pot_a2}) and (\ref{pot_n}), it  
can be instructive to use the classical description proposed by Landau for equilibrium phase transitions \cite{PB1989}. It has been shown, that this kind of description can be also obtained as a mean field approach for the Langevin equation of non-equilibrium systems with two $(Z_2)$ - symmetric absorbing states \cite{HCDM2005,VL2008}. 

In our paper we have written the master equation as a function of concentration $c = N_\uparrow/N$ of 'up'-spins. We have decided to use this quantity for convenience - calculations are simple and equations have compact forms. However, to meet the request of the symmetry \cite{HCDM2005,VL2008} one should use an order parameter (in this case magnetization) defined as:
\begin{equation}
\phi=\frac{N_\uparrow+N_\downarrow}{N}
\end{equation}
for which potentials (\ref{pot_a1}), (\ref{pot_a2}) and (\ref{pot_n}) are symmetric under reversal  $\phi \rightarrow -\phi$. 

Following the approach presented in \cite{HCDM2005,VL2008}, which coincides with the classical approach proposed by Landau, we expand potentials (\ref{pot_a1}), (\ref{pot_a2}) and (\ref{pot_n}), rewritten as a function of $\phi$,  into power series and keep only several terms of an expansion:
\begin{equation}
V(\phi)=A\phi^2+B\phi^4+C\phi^6,
\label{eq_Landau}
\end{equation}
where coefficients $A=A(p,q),B=B(p,q),C=C(p,q)$ depend on the model. 

For the model with independence:
\begin{eqnarray}
A(p,q)&=&-\frac{(1-p)}{2^q} \frac{(q-1)}{2} + \frac{p}{4}, \nonumber\\
B(p,q)&=&-\frac{(1-p)}{2^q}\frac{q(q-1)(q-5)}{24}, \nonumber \\
C(p,q)&=&-\frac{(1-p)}{2^q}\frac{q(q-1)(q-2)(q-3)(q-9)}{720}.
\label{eq_landau_indep}
\end{eqnarray}

From the Landau theory it is known that for $B(p,q)>0$ and $C(p,q)>0$ there is a critical point at which $A(p,q)$ changes sign \cite{PB1989}. For $A<0$ potential $V(\phi)$ has two symmetric minima and thus the system is driven to one of partially ordered state with $\phi \ne 0$. For $A>0$ potential $V(\phi)$ has minimum at $\phi=0$ and therefore the system remains in an active disordered state and a magnetization $\phi$ fluctuates around zero. From (\ref{eq_landau_indep}) it is easy to calculate that:
\begin{eqnarray}
A(p,q)=0 & \rightarrow & p=p^*=\frac{q-1}{q-1+2^{q-1}} \nonumber\\
A<0 & \rightarrow &  p<p^*,\; \phi \ne 0 \nonumber\\
A>0 & \rightarrow &  p>p^*,\; \phi = 0,
\end{eqnarray}
which coincides with the result (\ref{eq_crit_indep}) obtained from the not an approximate version of the potential (\ref{pot_n}).  

\begin{figure}
\begin{center}
\includegraphics[scale=0.4]{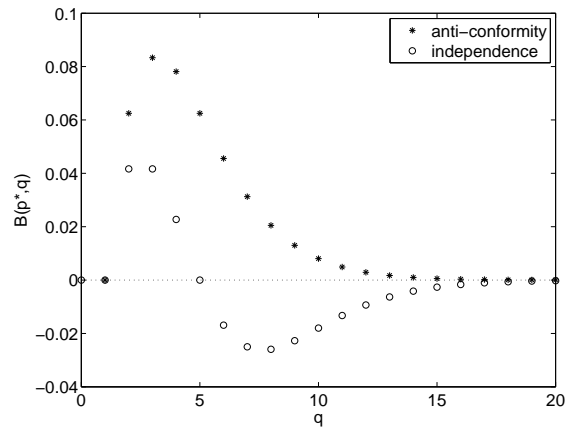}
\caption{Coefficient $B(p,q)$ (see an effective potential (\ref{eq_Landau})) for a critical point $p=p^*$ at which  $A(p,q)$ changes sign. For the model with independence (denoted by 'o') coefficient $B<0$ for $q>5$ which suggests discontinuous phase transition, whereas for the model with anticonformity (in the figure the model with anticonformity II is presented by '*') coefficient $B \ge 0$ for any value of $q$ and therefore the transition is continuous for arbitrary value of $q$.}
\label{fig_Bpc}
\end{center}
\end{figure}

As shown within the classical Landau theory, for $B(p,q)<0$ and $C(p,q)>0$ a discontinuous jump in the order parameter is expected \cite{PB1989}.  Again from (\ref{eq_landau_indep}) it is easy to see that $B(p,q)<0$ for $q>5$ (see also figure \ref{fig_Bpc}). Therefore we expect discontinuous phase transition for $q>5$, which also agrees with results obtained from (\ref{pot_n}). It should be mentioned here that transition for $B \le 0$ could possibly be included in the class of generalized voter model (so called unique GV transition) \cite{HCDM2005}. It has been noticed for a general class of models with two $(Z_2)$ - symmetric absorbing states that for $B \le 0$ the location of the potential minimum changes abruptly from $\phi=0$ to $\phi \pm 1$, i.e. discontinuous phase transition is observed \cite{HCDM2005}. In our model the situation is slightly different, because for $p>0$ there are no absorbing  states and below the transition point $|\phi|<1$. However, still the system jumps from a totally disordered to a partially  ordered state, i.e. discontinuous phase transition is observed. 

One should also notice that in the case of a q-voter model with independence for $q>9$ also $C(p,q)$ becomes negative and than the approximation (\ref{eq_Landau}) is not valid anymore. 

Analogous calculations can be done for the models with anti-conformity. Because both models with anti-conformity are qualitatively the same we present here results for the model with anti-conformity II. In this case: 
\begin{eqnarray}
A(p,q)&=& \frac{2pq-q+1}{2^{q}} \nonumber\\
B(p,q)&=&-\frac{1}{4}\frac{1-p}{2^{q}}\left[ {q-1 \choose 3} - {q-1 \choose 1} \right] \nonumber\\
&+& \frac{1}{4}\frac{p}{2^{q}}{q+1 \choose 3} \nonumber\\
C(p,q)&=&-\frac{1}{6}\frac{1-p}{2^{q}}\left[ {q-1 \choose 5} - {q-1 \choose 3} \right] \nonumber\\
&+& \frac{1}{6}\frac{p}{2^{q}}{q+1 \choose 5}.
\end{eqnarray}
Therefore in the case with anti-conformity:
\begin{eqnarray}
A(p,q)=0 & \rightarrow & p=p^*=\frac{q-1}{2q} \nonumber\\
A<0 & \rightarrow &  p<p^*,\; \phi \ne 0 \nonumber\\
A>0 & \rightarrow &  p>p^*,\; \phi = 0,
\end{eqnarray}
which coincides with the result (\ref{eq_crit_anti2}). 
Moreover, for the model with anti-conformity (see figure \ref{fig_Bpc}) coefficient $B(p=p^*,q) \ge 0$ for any value of $q$, i.e. the transition is continuous for arbitrary value of $q$.

\section{Conclusions}
\label{end}
In this paper we have asked the questions about the importance of the type of nonconformity (anti-conformity and independence) that is often introduced in models of opinion dynamics (see e.g. \cite{G2004,LLW2005,STT2011,NSC2012}). 
We realized that the differences between different types of nonconformity are very important from social point of view \cite{Nail_2000} but we have expected that they may be irrelevant in terms of microscopic models of opinion dynamics. 
To check our expectations we have decided to investigate nonlinear $q$-voter model on a complete graph, which has been recently introduced as a general model of opinion dynamics \cite{CMP2009}. 

To our surprise results for the model with anti-conformity are qualitatively different that for the model with nonconformity. In the first case there is a continuous order--disorder phase transition induced by the level of anti-conformity $p$. Critical value of $p$ grows with the size of the $q$-lobby. On the other hand for the model with nonconformity the value of the transition point $p^*$ decays with $q$. Moreover, the phase transition in this case is continuous only for $q \le 5$. For larger values of $q$ there is a discontinuous phase transition -- and coexistence of ordered (with majority) and disordered (without majority) phase is possible. 

We have suggested in the title and the introduction of the paper that both types of nonconformity play role of the noise. However, only independence introduces the real random noise that plays the similar role as a temperature. From this point of view the change of the type of transition reminds of the similar phenomena in the Potts model (for review see \cite{W82}). In the Potts model there is a first-order phase transition for $q>4$ and second-order phase transition for smaller values of $q$, where $q$ denotes the number of states of the spin. Of course in a case of our model $q$ does not denote the number of states, which is always 2, but the size of the group. In a way similar observation has been recently done by Araujo et. al. \cite{AAH2010} within the model of tactical voting. They have considered $q$ candidates on which citizens vote and proposed a balance function to quantify the degree of indecision in the society due to the coexistence of different opinions. It turned out that for some values of model parameters the model boiled down to the $q$-state Potts model, although similarly like in our model $q$ denoted the number of candidates instead of the number of states. The similar change of the type of transition has been also observed in a general class of systems with two $(Z_2)$ - symmetric absorbing states within a Langevin description \cite{HCDM2005,VL2008}. Moreover, it has been suggested that models with many intermediate states (i.e. the Potts model or a simple 3-state model described in \cite{VL2008}) behave as equivalent two-state models with effective transitions that are nonlinear in the local densities \cite{VL2008}, which is the case of $q$-voter model or a two-state model of competition between two languages \cite{AS2003}. The theory presented in \cite{HCDM2005,VL2008} suggests also that the continuous phase transition that is observed for $q<5$ could possibly be included in the Ising class, whereas discontinuous that is observed for $q>5$ in the class of generalized voter model.
 
Concluding the paper we would like to pay attention to one more phenomena that is visible in Fig. \ref{phase_tran}. For lobby $q=2$ results are the same for anti-conformity and independence. Therefore there is no surprise that the difference between two types of nonconformity has not been noticed while studding the Sznajd model ($i.e. q=2$) \cite{LLW2005,NSC2012}. 

\acknowledgements{The work was supported by funds from the National Science Centre (NCN) through grant no.
2011/01/B/ST3/00727.}

\end{document}